\documentclass[12pt,preprint]{aastex}

\newcommand{\Rmnum}[1]{\expandafter\@slowromancap\romannumeral #1@}

\shorttitle{Abrupt magnetic changes and UV emissions accompanying flares}
\shortauthors{Johnstone et al.}

\begin{document}


\title{Abrupt Longitudinal Magnetic Field Changes and Ultraviolet Emissions Accompanying Solar Flares}


\author{B.M. Johnstone, G.J.D. Petrie \& J.J. Sudol}
\affil{West Chester University, West Chester, PA 19383\\
  National Solar Observatory, Tucson, AZ 85719, USA}



\begin{abstract}
We have used \textit{Transition Region and Coronal Explorer} (TRACE) 1600~\AA\ images and \textit{Global Oscillation Network Group} (GONG) magnetograms to compare ultraviolet (UV) emissions from the chromosphere to longitudinal magnetic field changes in the photosphere during four X-class solar flares. An abrupt, significant, and persistent change in the magnetic field occurred across more than ten pixels in the GONG magnetograms for each flare. {These magnetic changes lagged the GOES flare start times in all cases, showing that they were consequences and not causes of the flares.} Ultraviolet emissions were spatially coincident with the field changes. The UV emissions tended to lag the GOES start times for the flares, and led the changes in the magnetic field in all pixels except one. The UV emissions led the photospheric field changes by 4 minutes on average with the longest lead being 9 minutes, however, the UV emissions continued for tens of minutes, and more than an hour in some cases, after the field changes were complete. The observations are consistent with the picture in which an Alfv\'{e}n wave from the field reconnection site in the corona propagates field changes outward in all directions near the onset of the impulsive phase, including downwards through the chromosphere and into the photosphere, causing the photospheric field changes, whereas the chromosphere emits in the UV in the form of flare kernels, ribbons and sequential chromospheric brightenings during all phases of the flare.
\end{abstract}



\section{Introduction}
\indent 

magnetohydrodynamics: Sun, solar magnetic fields, solar photosphere, solar chromosphere, EUV

Solar flares are believed to be caused by a restructuring of the solar coronal magnetic field, releasing energies up to around $10^{32}$~erg (Priest~1982). The most plausible source of flare energy is understood to be free magnetic energy built up in the coronal field by the turbulent photosphere over time, stored in the form of Maxwell stresses. While in the photosphere the plasma dominates the magnetic field, in the low corona the field dominates the plasma so that the Lorentz force is not opposed by a significant plasma force and is therefore small.  The coronal plasma is very highly conducting, preventing the coronal field from releasing its free energy in a gradual way, so the stresses continue to build until the energy is abruptly released as the coronal field suddenly rearranges itself in a simpler, less stressed configuration. Hudson~(2000), Hudson, Fisher and Welsch~(2008) and Fisher et al.~(2012) argue that after a coronal magnetic eruption, the remaining coronal field must implode in general, contracting downward, resulting in the field becoming more horizontal in the photospheric layer. Alfv\'{e}n waves or MHD fast-mode waves in the corona, resulting from the collapse of coronal post-flare loops, propagate toward the photosphere, and could carry enough energy to the photosphere to change the magnetic field there (Hudson, Fisher and Welsch~2008). Such a wave would cause both ultraviolet (UV) and H$\alpha$ emissions in the chromosphere as well as the magnetic field changes in the photosphere. The relationship between measured changes in the photospheric field and chromospheric emission is the subject of this paper.

Wang et al.~(1992, 1994) found rapid and permanent field changes in flaring active regions, but a number of later studies produced inconclusive results. In recent years the evidence for photospheric field changes during flares has steadily increased. Wang and Liu~(2010) studied 11 X-class flares for which vector magnetograms were available, and found in each case an increase of transverse field at the polarity inversion line. The HMI vector data for two major flares from NOAA~11158 have been studied in several papers using a variety of methods (Wang et al.~2012, Gosain~2012, Sun et al.~2012, Liu et al.~2012, Petrie~2012), yielding results consistent with Hudson, Fisher and Welsch's~(2008) loop-collapse scenario. Following the work of Sudol and Harvey~(2005), Petrie and Sudol~(2010) used one-minute GONG longitudinal (line-of-sight) magnetograms to characterize the spatial distribution, strength and rate of change of permanent field changes associated with 77 flares of GOES class at least M5 and found statistically significant correlations in the field changes consistent with  the loop-collapse scenario. Burtseva and Petrie~(2012) refined the magnetic flux calculations and confirmed the results of Petrie and Sudol~(2010) using a feature-tracking analysis of the GONG data. In a combined analysis of GOES X-ray emission data and GONG magnetograms, Cliver and Petrie~(2012) found a sharp change in the magnetic flux coincident with the onset of the flare impulsive phase and a correspondence between the end of the stepwise change and the time of peak SXR emission. They identified the abrupt changes in photospheric magnetic fields as an impulsive phase phenomenon and indicated that the coronal magnetic field changes that drive flares are rapidly transmitted to the photosphere.

The morphology of a flare in the lower atmosphere can be used to trace the progress of coronal magnetic energy release. H$\alpha$ and UV observations of solar flares, interpreted as foot-points of flaring loops, have been used to infer the magnetic energy release during the flares. The expanding flare ribbons in H$\alpha$ and UV are the signature of ongoing magnetic reconnection in the corona as fields reconnect at ever higher altitudes (Qiu et al.~2002). Past theoretical and observational work (Forbes and Lin~2000, Zhang et al.~2001, Qiu et al.~2002, 2004, Cheng et al.~2003) has found a correlation between the magnetic reconnection rate and X-ray flux. Besides flare kernels and ribbons, sequential chromospheric brightenings (SCBs) have been associated with flares (Balasubramaniam et al.~2005) and interpreted as a signature of chromospheric evaporation associated with tether-cutting reconnection during eruptions (Pevtsov et al.~2007). Whereas flare kernels and ribbons are only observed during flares, SCBs can occur before, during and after flares.

The spatio-temporal relationship between photospheric field changes and lower atmospheric brightenings has not been investigated in detail. To address this issue we will analyze in this paper four X-class solar flares using \textit{Transition Region and Coronal Explorer} (TRACE, Handy et al.~1999) 1600~\AA\ images and one-minute \textit{Global Oscillation Network Group} (GONG) longitudinal magnetograms. Sudol and Harvey~(2005) compared the UV emissions to magnetic field changes for three out of four of the flares in our analysis. They analyzed a total of 14 representative pixels located in each distinct area in an active region where a field change occurred. They observed that the magnetic field changes and UV brightenings correlated to within $0\,^{\circ}.5$ in heliographic coordinates, and to within 4 minutes, and suggested that these flare phenomena are spatially and temporally coincident. Sudol and Harvey also reported that a UV emission always accompanied a magnetic field change. Our analysis differs from theirs in that we will examine individual pixels from the entire active region and seek to quantify the time delay.

The paper is organized as follows. We describe the data and data analysis in Sections~\ref{sect:data} and \ref{sect:dataanalysis}. We summarize the statistics of the magnetic changes and UV brightenings in Section~\ref{sect:statanalysis} and conclude with a discussion in Section~\ref{sect:conclusion}.

\section{Data}
\label{sect:data}

\begin{table}
\begin{center}
\caption{List of Flares in This Survey}\
\label{flarelist}
\\
\begin{tabular}{lcccr} \hline\hline
Date(UT) & GOES Times           & GOES class & Location & NOAA  \\ 
               & Start/Peak/End    &                   &                & Number \\ \hline
2001 August 25$^{\dagger}$ & 1623/1645/1704 & X5.3 & S17E28 & 09591 \\
2003 October 26$^{\dagger}$ & 0557/0654/0733 & X1.2 & S17E39 & 10487 \\
2003 October 29$^{\dagger}$ & 2037/2049/2101 & X10.0 & S16W09 & 10486 \\
2004 July 16 & 1349/1355/1401 & X3.6 & S10E34 & 10649 \\
\end{tabular}    
\end{center}
\hspace{60pt}$^{\dagger}$Previously analyzed by Sudol and Harvey~(2005)
\end{table}

\indent Changes in the magnetic field occur on a timescale of 10 minutes. Therefore, in order to distinguish the field changes associated with a flare, we chose flares for which continuous GONG one-minute data were available for at least an hour before and after the flare. We reduced the initial group of flares observed by the GONG telescopes by excluding flares with an apparent central meridian longitude difference greater than $65\,^{\circ}$ and flares categorized below M5. This smaller group was further reduced to four by eliminating flares that did not have continuous TRACE 1600~\AA\ data available of comparable cadence to the GONG data. The four flares analyzed here, all X-class flares, are listed in Table \ref{flarelist}.

GONG has six sites stationed throughout the world that produce full-disk images of the relative Doppler shift of the Ni~{\sc I} line at 676.8~nm at a one minute cadence. The cadence of the TRACE 1600\AA\ images is non-uniform but on the order of two minutes. GONG magnetograms have 860x860 pixels and have a field of view of 36'x36' whereas TRACE images have 768x768 pixels and have a field of view of 8.5'x8.5'. 

We used magnetograms covering the four-hour interval centered on the GOES soft X-ray start time for each flare, and TRACE images covering from up to 30 minutes before and 40 minutes after the start time for each flare, with additional images at 90 and 120 minutes.

\section{Data Analysis}
\label{sect:dataanalysis}

\indent We used the same data analysis method as Sudol and Harvey~(2005). The GONG magnetograms were converted from velocities to magnetic field strength with the conversion factor 0.352 G \(m^{-1}s^{-1}\). The full-disk GONG magnetograms were remapped to an overhead, azimuthal-equidistant projection, $32\,^{\circ}$x$32\,^{\circ}$ field of view in heliographic coordinates (256x256 pixels), tangent to a point near the center of the flaring active region. Each remapped magnetogram was registered to a reference image which was composed of an average of the ten images obtained prior to the GOES X-ray start time of the flare. Tests of the motion of the centers of sharp features in the GONG magnetograms suggest that there is less than one pixel error associated with the registration which is due to uncertainty in the direction of true north and the natural motion of surface features. The TRACE images were remapped to the same coordinates and scale as the GONG magnetograms. The remaps were then registered to the image taken prior to the start of the flare. 

\begin{figure}[h]
\begin{center}
\resizebox{0.75\hsize}{!}{\includegraphics*{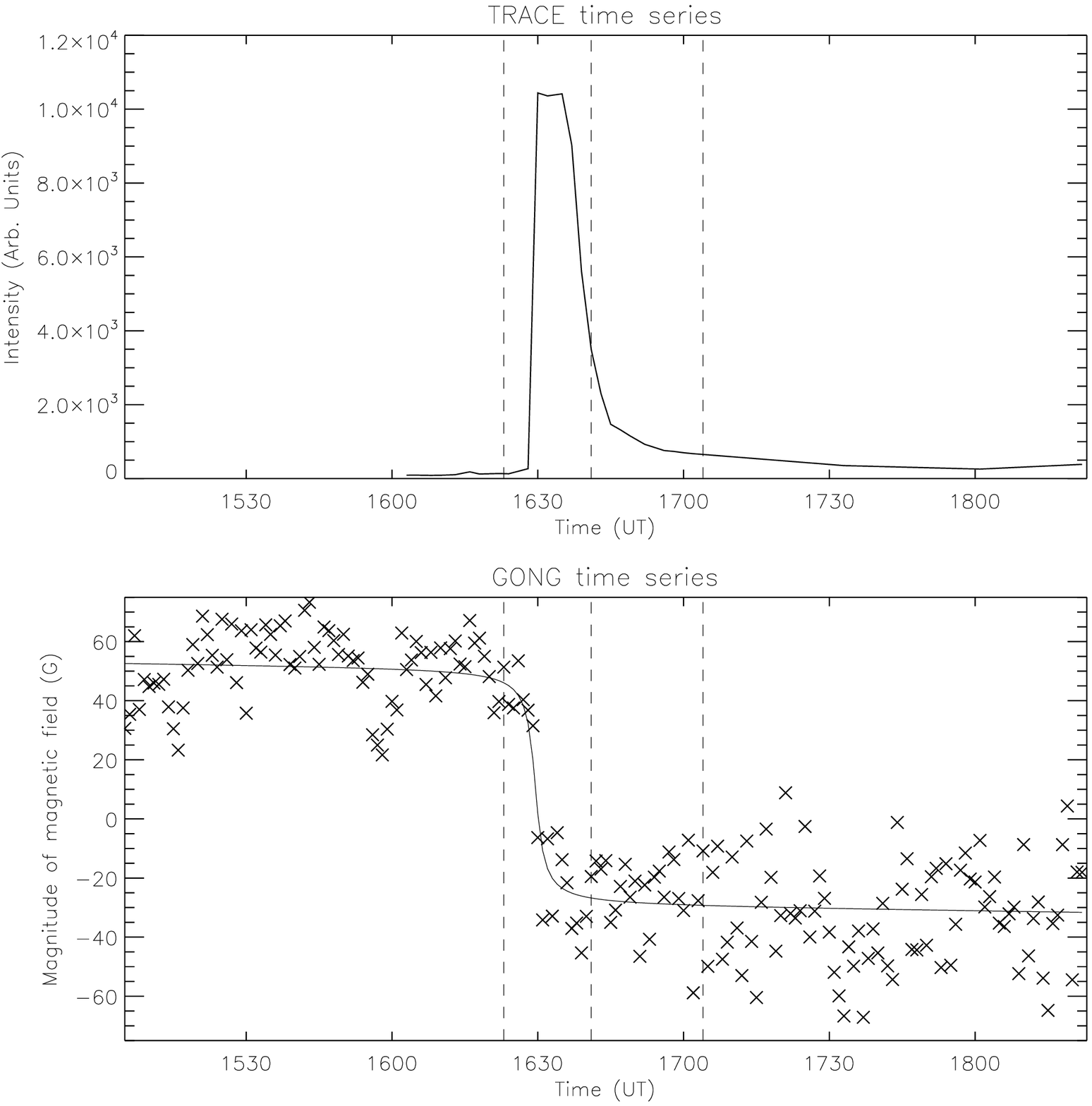}}
\end{center}
\caption{\small Shown is the TRACE (top) and GONG (bottom) time series for a pixel from the flare that occurred on 2001 August 25. Equation~\ref{atancurve} has been fitted to the time series of measurements from the GONG pixel. The GOES X-ray start, peak and end times are indicated by vertical dotted lines in both plots. The size of the step, $2c$, for this particular pixel is approximately 85 G and the step has a duration of 6 minutes.}
\label{timeseries}
\end{figure}

Field changes occur at different times and with different amplitudes and durations at different locations across a flaring region. We determine the spatio-temporal distribution of the field changes by analyzing each pixel's time series individually. This approach allowed Sudol \& Harvey~(2005) to investigate the relationship of propagating patterns of field change to motions of H$\alpha$ flare ribbons. Here it enables us to study the relationship in space and time between field changes and EUV brightenings. A time series plot was constructed for each pixel in the GONG magnetograms. The following step function was fit to each of the time series plots, 
\begin{equation}
B(t)=a+bt+c\left\{ 1+\frac{2}{\pi}tan^{-1}[n(t-t_{0})]\right\} ,
\label{atancurve}
\end{equation}
\noindent where $a$ and $b$ account for the strength and evolution of the background field, $c$ represents the half-amplitude of the field change, $n$ represents the inverse of the amount of time it took the field change to occur, and $t_{0}$ is the time corresponding to the midpoint in the step function. The amplitude of the step, $2c$, equals the change in the magnetic field, $dB$, and $\mathrm{d}t = \frac{\pi}{n}$ is an estimate for the time interval of the change in the magnetic field. An example of a time-series and the fit to that time series from the 2011 August 25 X5.3 flare appears in Figure~\ref{timeseries}. GONG and TRACE images of this flare are shown in Sudol and Harvey's~(2005) Figure~11.

The time series were initially required to pass a goodness-of-fit test: the size of the step had to be at least 1.4 times the scatter in the data prior to the flare with respect to the fit of Equation~(\ref{atancurve}) to the data. The pixels that passed this automated test were then examined by eye. Sudol and Harvey (2005) found that the amplitudes of the field changes were between 1.4 times and 20 times the standard deviation of the noise in the preflare time series. Sudol \& Harvey looked at 10,000 light curves and selected the ones they were confident showed a step-like field change. The lowest field changes had amplitudes of size 1.4 times the noise in that sample. Petrie \& Sudol (2010) found similar results. The threshold value of 1.4 was found to be optimal in excluding insignificant changes and artifacts from consideration. Adjustments to that number in the software result in either too few positives or too many negatives. The quantitative selection criteria reduced $860\times 860$ pixels per flare to a few hundred in which field changes might have occurred. Visual inspection followed to eliminate false positives. No known computer algorithm can match the eye at edge detection, but reviewing millions of pixels by eye is taxing, so our data processing made use of both a computer algorithm and the human eye. The criteria that we used to eliminate cases in which no significant field change occurs, which is typical for more than 99\% of the pixels in a field, are conservative in that they allow through many false signatures of field change. We then reviewed by eye all of the pixels that met our criteria to eliminate the false positives. Sudol \& Harvey~(2005) discuss this aspect of the data processing in detail on pp. 650-652. 

In all, 77 pixels met our criteria and passed our visual review. For each selected pixel from the GONG magnetograms, we constructed a light curve for the corresponding pixel in the remapped TRACE images.

The following expression was used to establish the start time for the change in the magnetic field,
\begin{equation}
t_{start}=t_{0}-\frac{\pi}{2n}.
\label{GONGstarttime}
\end{equation}
\noindent Since $dt$ is the time interval of the field change, half of $dt$ from the time corresponding to the midpoint of the step function results in the start time. To determine the start time for the UV brightening, a line was fitted to the leading edge of the light curve. UV light curves showed a sharp leading edge, so the start time was taken to be the point at which the line intersected the background. We took the cadence of the GONG magnetograms, one minute, as a conservative estimate of the uncertainty in the start time for the magnetic field changes. Similarly, we took the average cadence of the 1600~\AA\ TRACE images in our data set, 2 minutes, as the uncertainty in the start time of the UV brightening. 

\begin{figure}[h]
\begin{center}
\resizebox{0.33\hsize}{!}{\includegraphics*{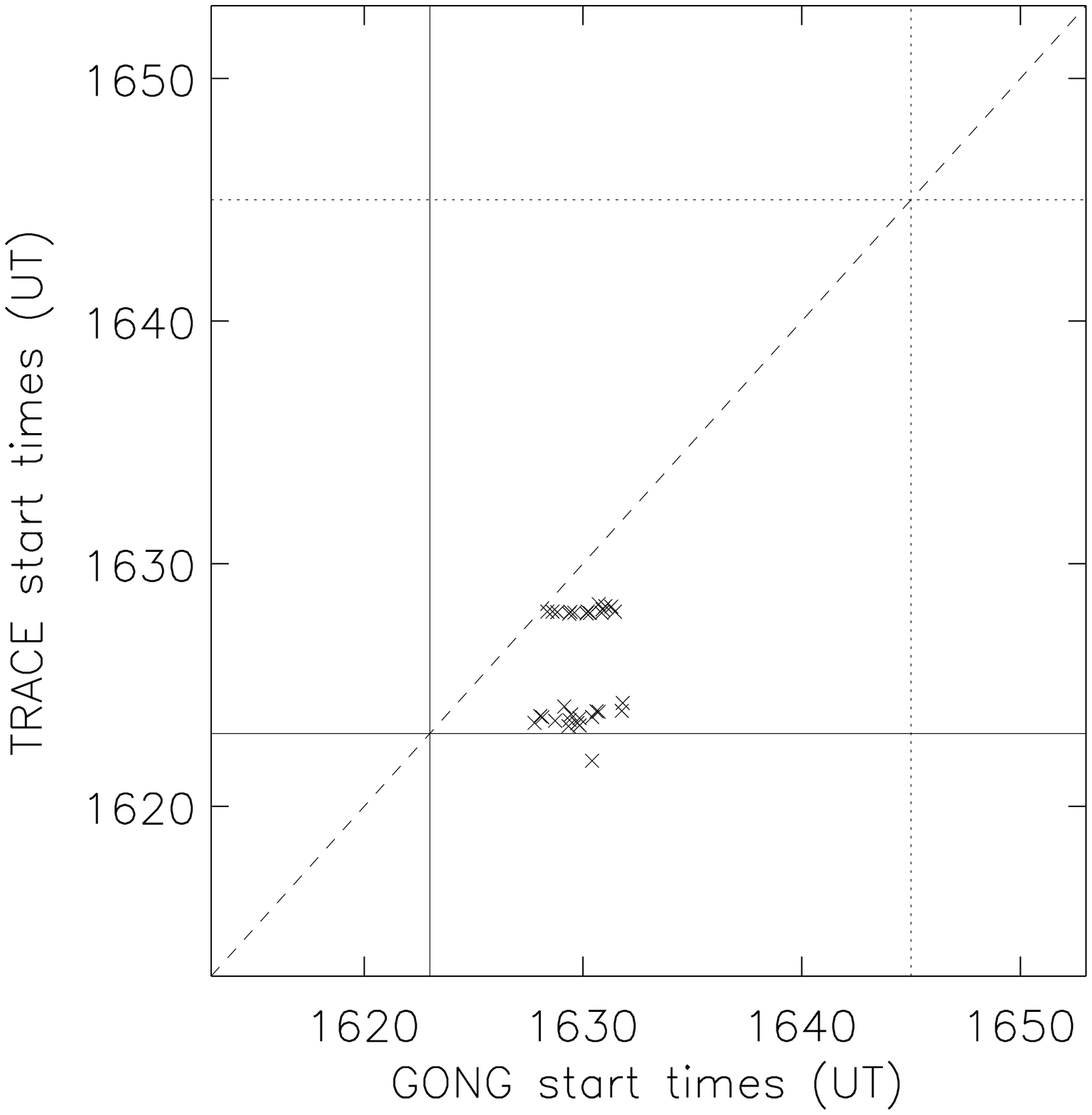}}
\resizebox{0.33\hsize}{!}{\includegraphics*{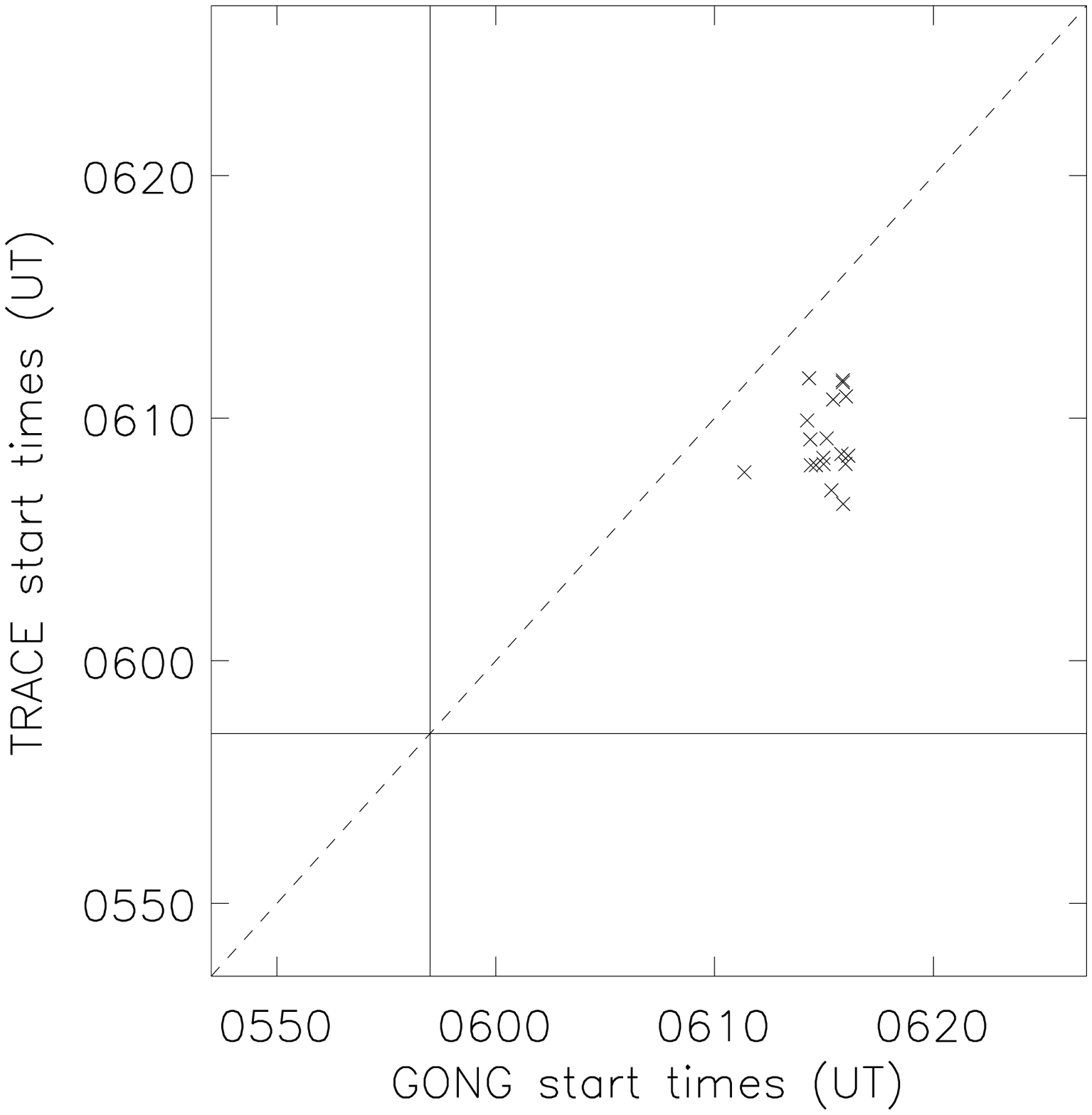}}
\resizebox{0.33\hsize}{!}{\includegraphics*{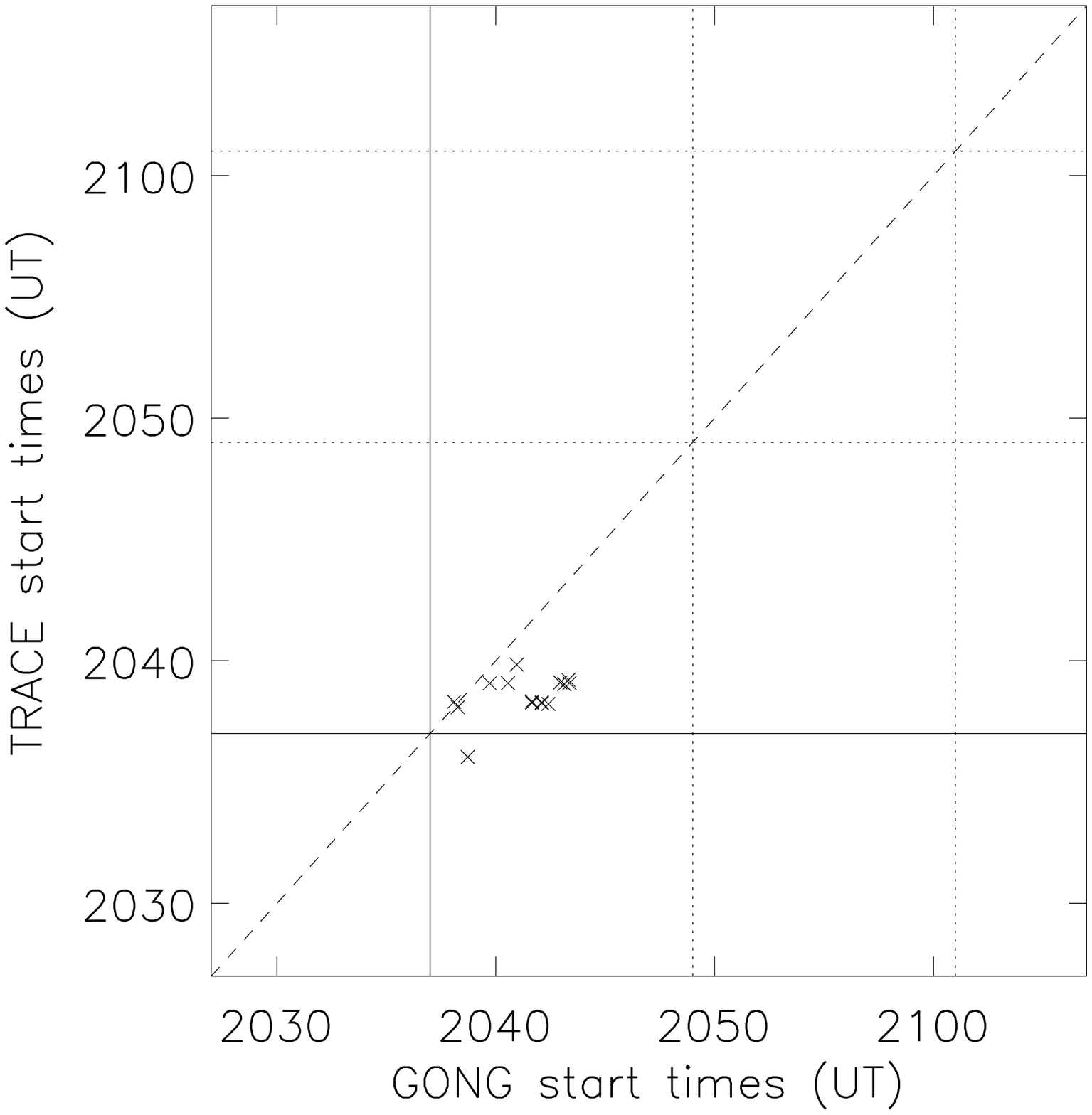}}
\resizebox{0.33\hsize}{!}{\includegraphics*{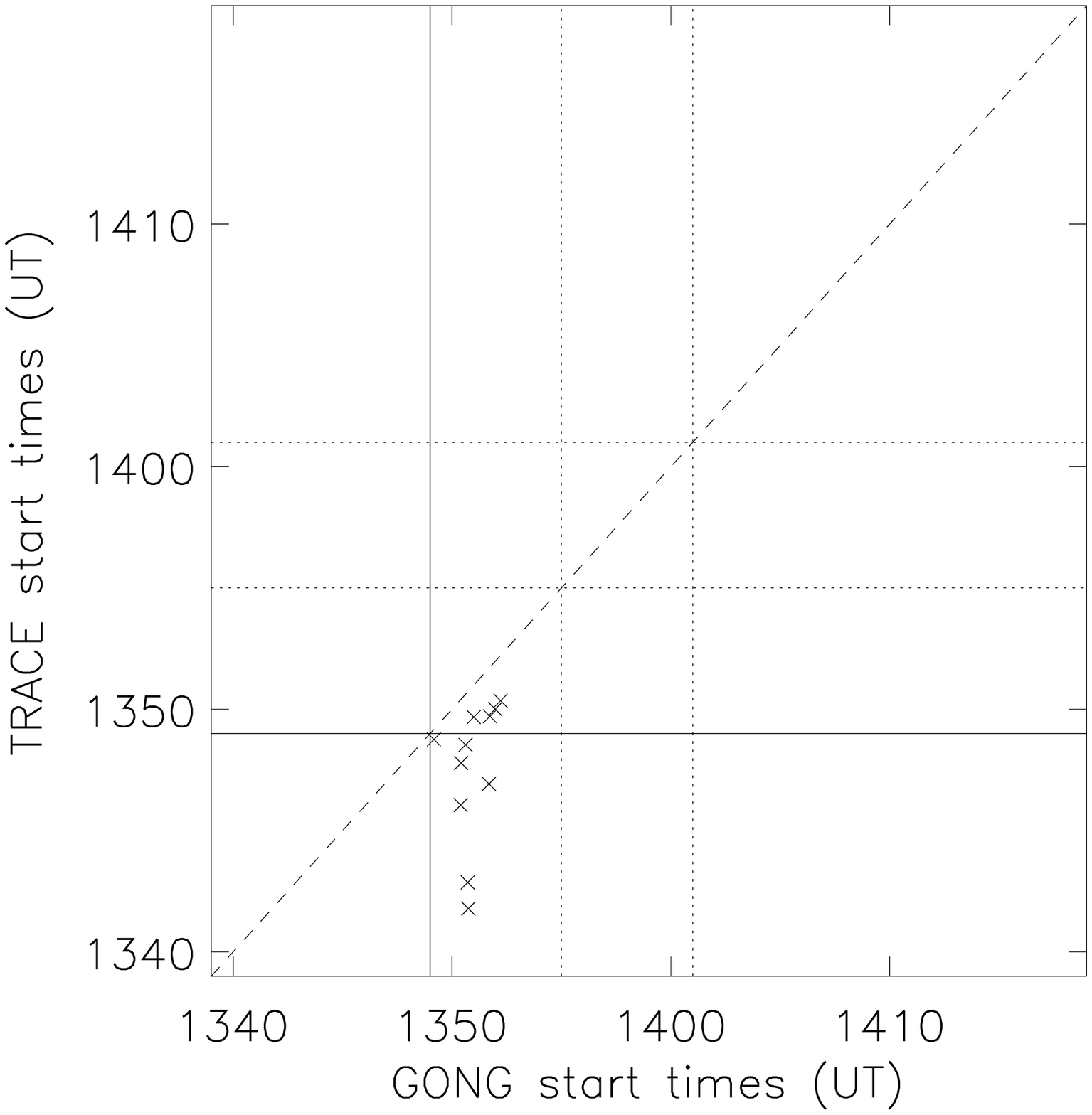}}
\end{center}
\caption{Scatter plots of TRACE and GONG start times for the 2001 August 25 X5.3 (top left), 2003 October 26 X.2 (top right), 2003 October 29 X10.0 (bottom left) and 2004 July 16 X3.6 (bottom right) flares. Each data point represents the same pixel in the GONG and TRACE images. The horizontal and vertical solid lines represent the GOES start time for the flare and the dotted lines represent the GOES peak and end times, not all of which fall within the range of the plot - see Table~\ref{flarelist}. For pixels above the horizontal axis the TRACE brightenings began after the GOES SXR flux. For pixels to the right of the vertical axis the GONG field changes began after the published GOES X-ray flare start time. For pixels below the oblique dashed line, which represents where coincident events would occur, the GONG field changes lagged the TRACE UV emissions.}
\label{starttimes}
\end{figure}

\begin{figure}[h]
\begin{center}
\resizebox{0.33\hsize}{!}{\includegraphics*{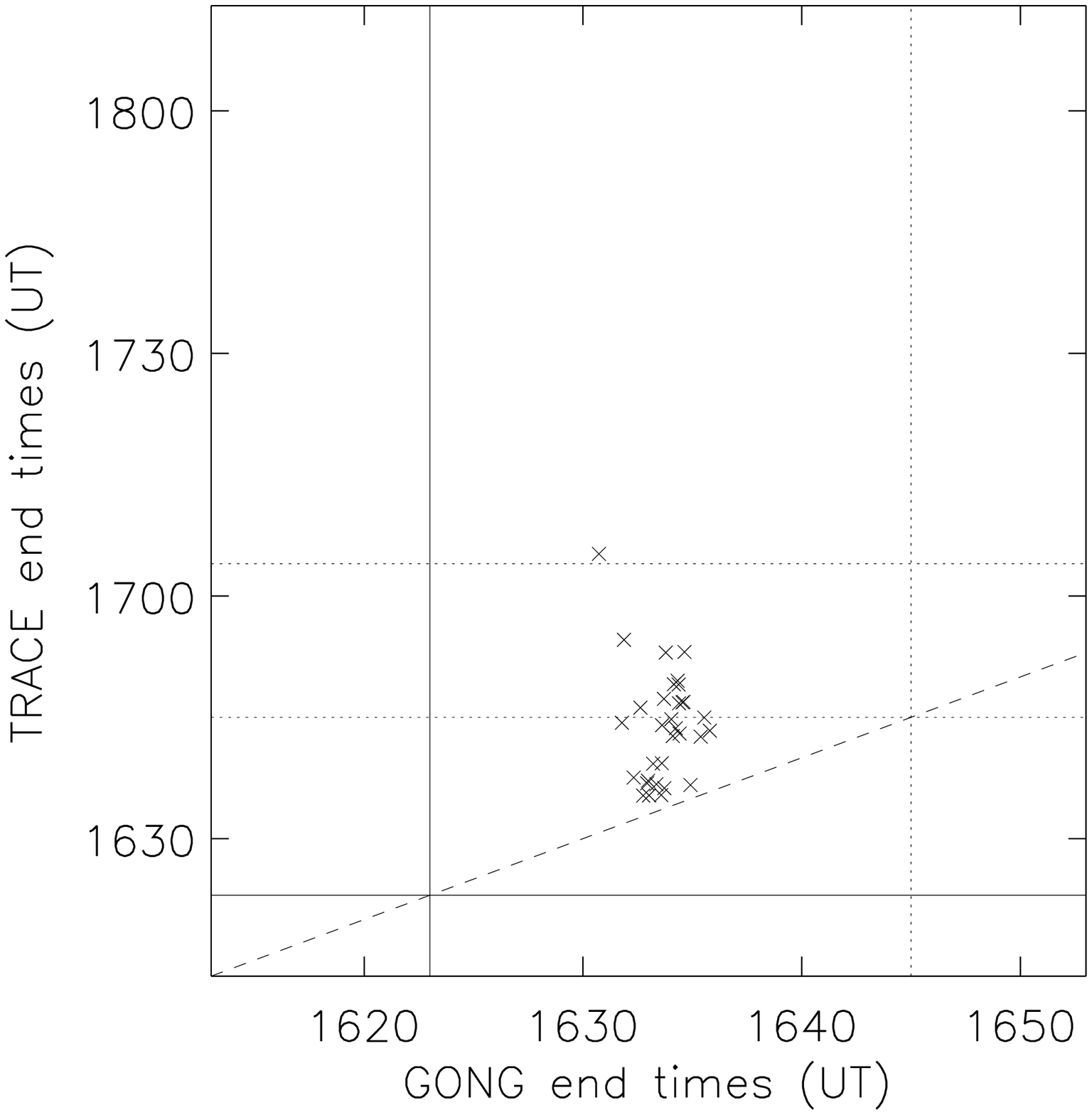}}
\resizebox{0.33\hsize}{!}{\includegraphics*{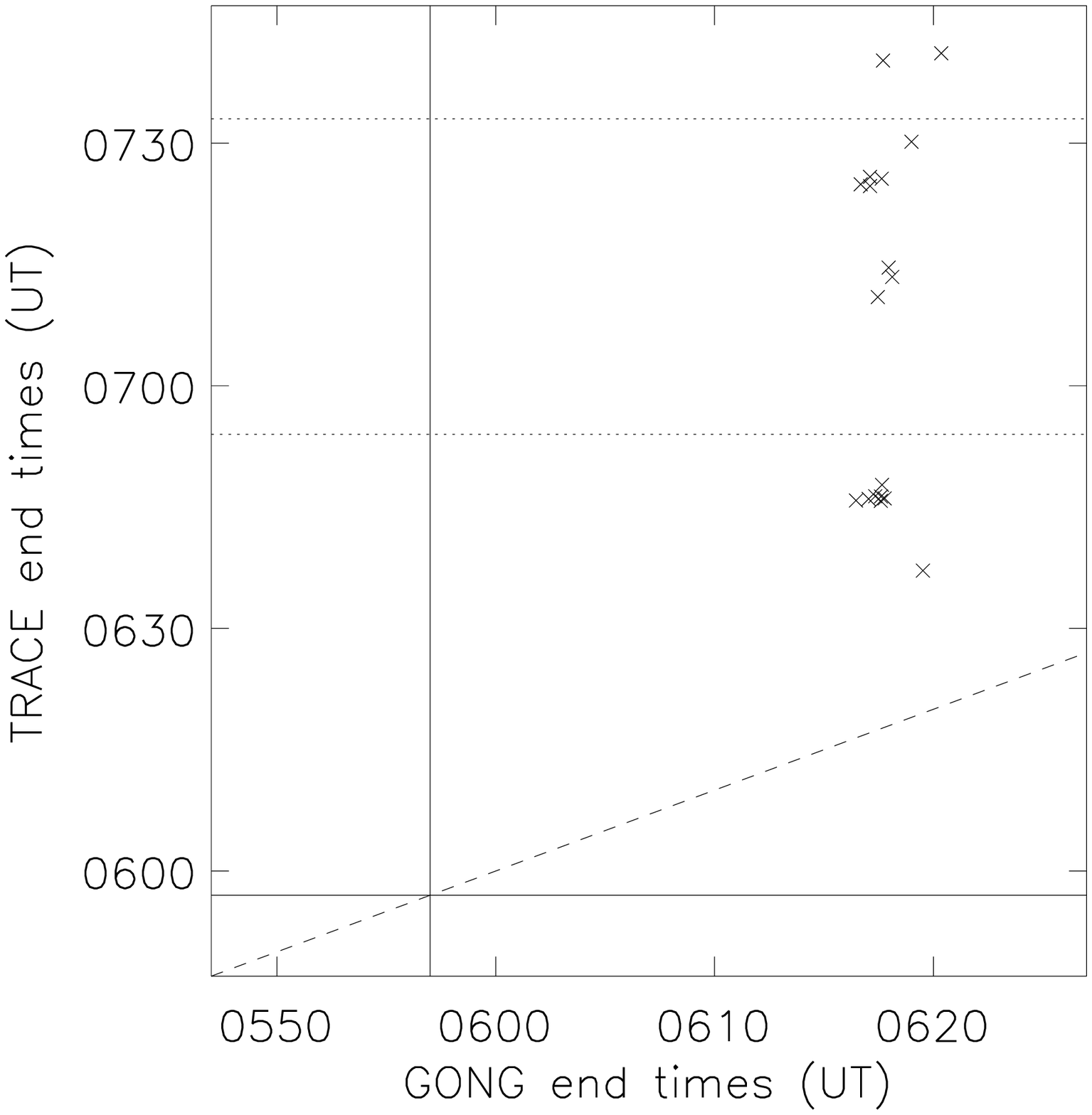}}
\resizebox{0.33\hsize}{!}{\includegraphics*{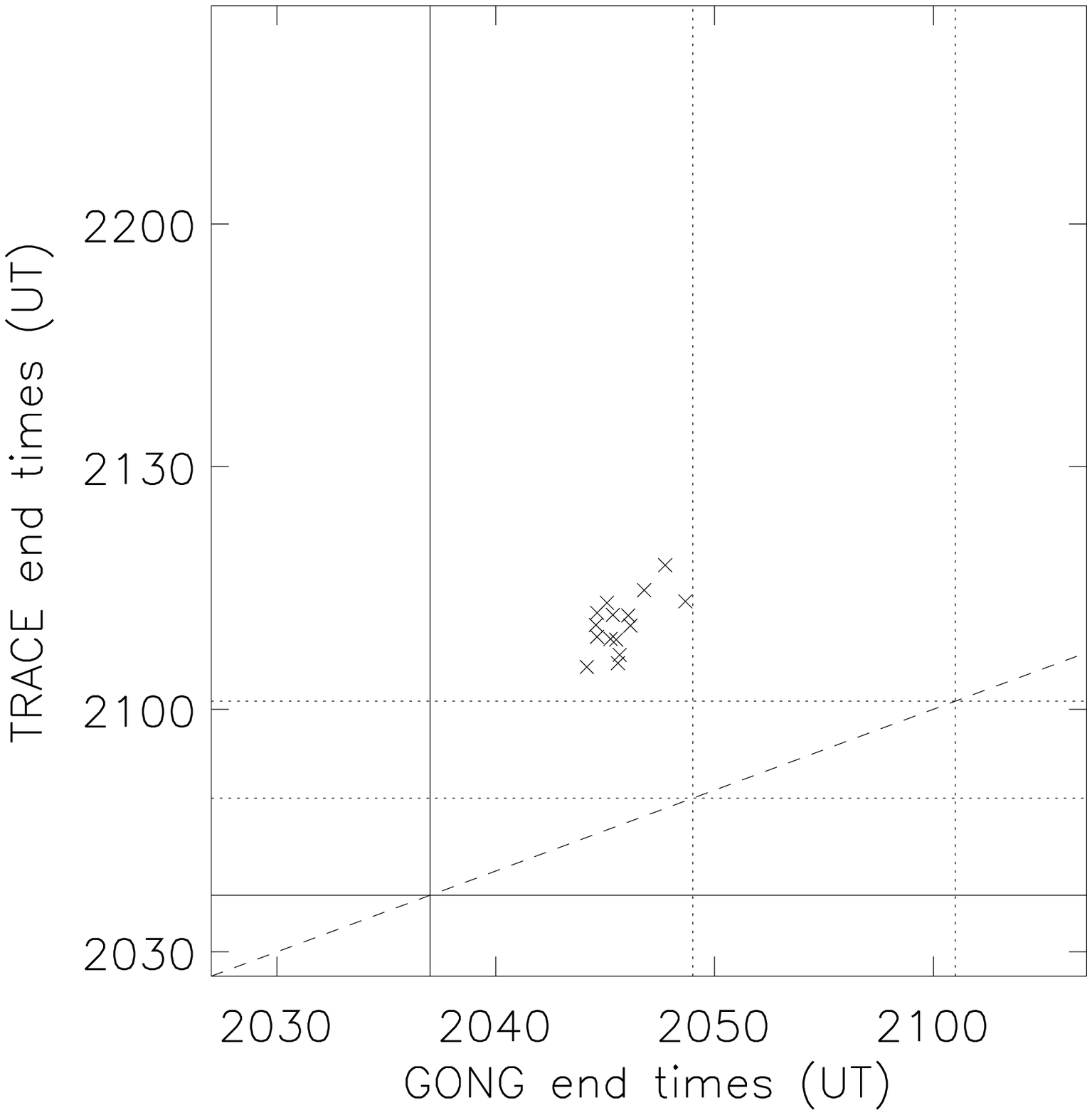}}
\resizebox{0.33\hsize}{!}{\includegraphics*{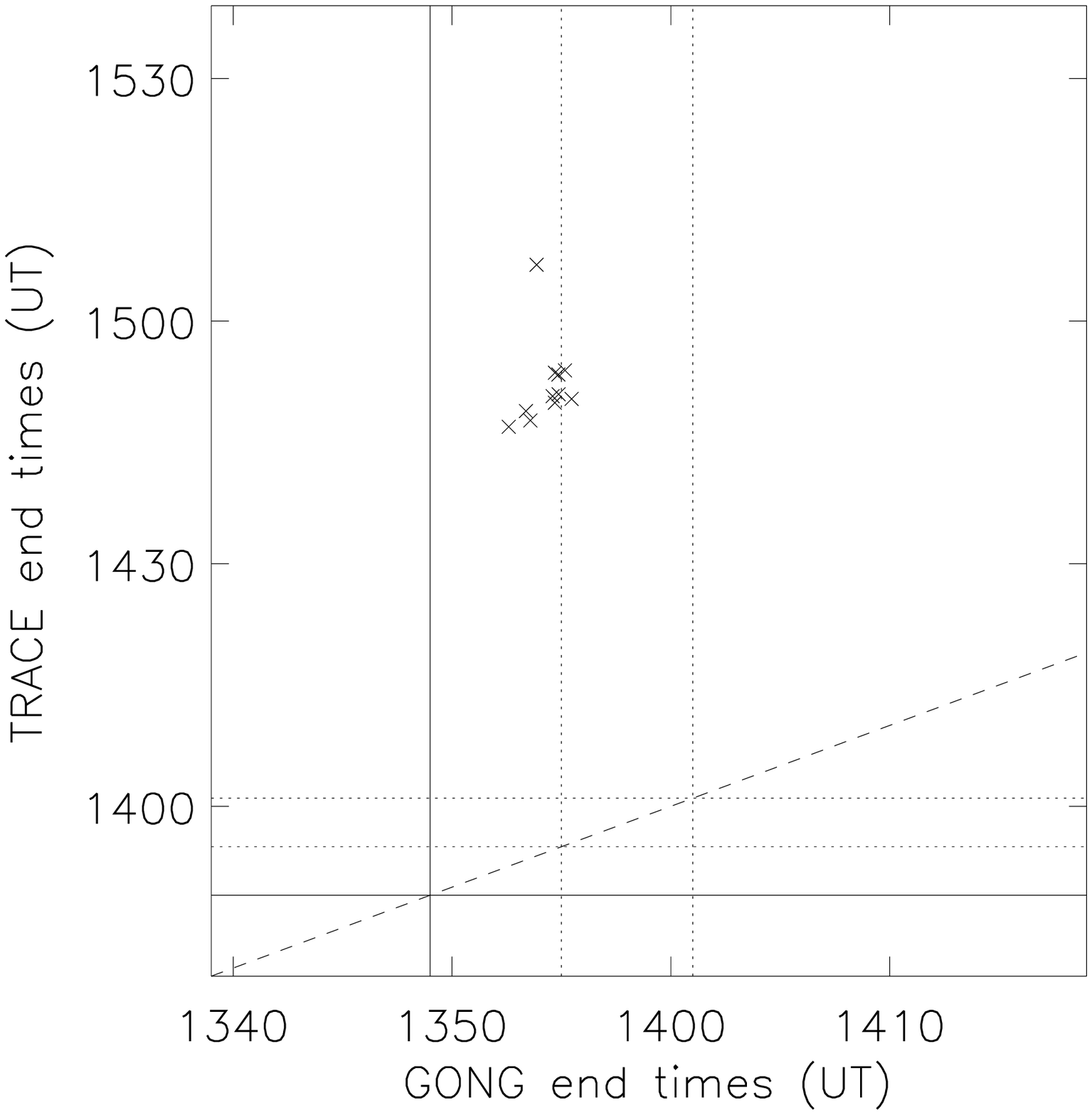}}
\end{center}
\caption{Scatter plots of TRACE and GONG end times for the 2001 August 25 X5.3 (top left), 2003 October 26 X.2 (top right), 2003 October 29 X10.0 (bottom left) and 2004 July 16 X3.6 (bottom right) flares. Each data point represents the same pixel in the GONG and TRACE images. The horizontal and vertical solid lines represent the GOES start time for the flare and the dotted lines represent the GOES peak and end times, not all of which fall within the range of the plot  - see Table~\ref{flarelist}. For pixels above the oblique dashed line, which represents where coincident events would occur, the GONG field changes ended before the TRACE UV emissions.}
\label{endtimes}
\end{figure}

\section{Statistical analysis}
\label{sect:statanalysis}

For each of the four flares, we found at least ten pixels in the active region where the longitudinal magnetic field underwent an abrupt, permanent and significant stepwise change. We found 77 such pixels in total. In terms of the fit parameters, abrupt means the magnetic field change was complete in less than 10 minutes. Permanent means that the field change persisted until the end of our data window. Significant means that the amplitude of the field change was greater than 1.4 times the scatter of the data with respect to the fit prior to the field change (see Section~\ref{sect:dataanalysis}). These pixels do not themselves cover the majority of the flaring region areas but they do generally represent the major field changes in the region. Petrie and Sudol~(2010) found that individual pixels with the most abrupt and significant changes give results consistent with calculations based on total flux changes. Petrie and Sudol's results have recently been confirmed by Burtseva and Petrie~(2012) using a feature tracking technique. The changes tend to occur in clusters of pixels with similar behavior, with the chosen pixels exhibiting the largest, most abrupt changes of their clusters.

The ratio of magnetic field increases to decreases was consistent with Sudol and Harvey's~(2005) and Petrie and Sudol's~(2010) results. The average magnitude of the magnetic field change was approximately 100 G, however, field changes ranged from 20~G to 300~G. We found, as did Sudol and Harvey~(2005), that a chromospheric UV brightening accompanied every pixel which underwent an abrupt magnetic field change, however, not every brightening was accompanied by a detected field change. The chromospheric brightenings also lasted longer than the field changes in general. Figures~\ref{starttimes} and \ref{endtimes} and Table~\ref{avgtime} give details of the timings.

Figure~\ref{starttimes} shows that the field changes in all 77 GONG pixels began after the published GOES X-ray flare start times, with the delays ranging from 1 minute to 19 minutes. This observation supports the theoretical interpretation in which the coronal event causes the photospheric change and not vice versa. Sudol and Harvey~(2005) were the first to make this point and Cliver and Petrie~(2012) recently reinforced this result in a detailed comparative analysis of the GONG and GOES data. All of the field changes, however, occurred before the GOES peak times, a result consistent with the correspondence between the end of the stepwise change and the time of peak SXR emission found by Cliver and Petrie~(2012).

On the other hand, for three of the flares a minority of TRACE pixels brightened significantly before the published GOES times. These early UV brightenings may be associated with SCBs (Balasubramaniam et al.~2005). In all pixels, the chromospheric brightening began before the photospheric field changes.

From Figure~\ref{endtimes} it is clear that the TRACE UV brightenings continued after the GONG field changes were complete in every pixel, an hour or more in some cases. The GONG field changes were complete before or around the GOES peak times and significantly before the GOES end times. This is in line with the findings of Cliver and Petrie~(2012). The TRACE UV brightenings continued after even the GOES end times in some cases.

\begin{table}
\begin{center}
\caption{Differences between GONG and TRACE start and end times in minutes. Times are positive if GONG lags TRACE. Errors are standard deviations.}\
\label{avgtime}
\\
\begin{tabular}{lcccc} \hline\hline
Flare    & Start Time  & End Time & GONG & TRACE\\ 
(UT Date)                & Delay    & Delay & Duration & Duration\\ \hline
2001 August 25 & $4 \pm 2$ & $-10 \pm 7$ & $4 \pm 2$ & $18 \pm 17$ \\
2003 October 26 & $6 \pm 2$ & $-50 \pm 22$ & $3 \pm 2$ & $58 \pm 21$\\
2003 October 29 & $3 \pm 2$ & $-25 \pm 3$ & $4 \pm 2$ & $32 \pm 3$\\
2004 July 16 & $3 \pm 3$ & $-58 \pm 5$ & $3 \pm 1$ & $65 \pm 8$\\ 
All & $4 \pm 2$ & $-29 \pm 23$ & $4 \pm 2$ & $37 \pm 23$\\
\end{tabular}    
\end{center}
\end{table}

The average start time delays, end time delays and durations of the UV emission and the magnetic field change appear in Table~\ref{avgtime}. The average time delay between the start of the UV emission and the start of the change in the magnetic field was 4$\pm$2 minutes, with the longest time delay being approximately 9 minutes. The average time delays from one flare to the next show some differentiation - see Table~\ref{avgtime}. In all but one pixel the UV emission started before the change in the magnetic field. In all of the pixels, the UV emission ended after the change in the magnetic field as is clear from Table \ref{avgtime}. The average time delay between the end of the change in the magnetic field and the end of the UV emission was 29$\pm$22 minutes, with the longest time delay being approximately 83 minutes. Table~\ref{avgtime} shows that there was considerably more spread in the end time delays than in the start time delays, but the patterns are clear. The average TRACE duration was 37 minutes and the average GONG duration was 4 minutes. TRACE durations ranged between 7 and 93 minutes whereas GONG durations ranged between 1 and 19 minutes.

\section{Conclusion}
\label{sect:conclusion}

Sudol and Harvey~(2005) suggested that the UV brightenings and the change in the magnetic field associated with flares may be both spatially and temporally coincident. Comparing UV emissions in the chromosphere to magnetic field changes in the photosphere using TRACE 1600~\AA\ images and GONG magnetograms for four X-class solar flares, we found that they were spatially coincident but their temporal relationship is more complex. The chromospheric emissions tended to lead the photospheric field changes by a few minutes and continue tens of minutes after the field changes are complete. There is more spread in the gaps between the photospheric and chromospheric end times than between the chromospheric and photospheric start times but the patterns are statistically significant. 

The magnetic changes lagged the GOES start times in every pixel. This is in line with the results of Sudol and Harvey~(2005) and Cliver and Petrie~(2012) and it shows that the flares caused the photospheric changes and not vice versa. We now need to explain why the chromospheric brightenings tended to begin before the magnetic changes and to last much longer than them.

Sudden magnetic field changes caused by flares would be transmitted from the coronal flare location to the lower atmospheric layers by a fast MHD wave or an Alfv\'en wave. Because there is a large change in the magnetic and plasma parameters over a short distance between the base of the corona and the photosphere, much of the energy of the wave will be reflected at this transition layer. Emslie \& Sturrock~(1982) estimate the fraction of energy transmitted to the lower atmosphere $T_E$ as $T_E = 4\Theta^{1/2} / ( \Theta^{1/2} +1)^2$ where $\Theta$ is the ratio between the coronal and lower atmospheric temperatures. For a wave reaching the chromosphere they estimate that $T_E$ is between about 0.7 and 0.3, taking $\Theta\approx 10-100$. For the photosphere, Hudson, Fisher and Welsch~(2008) adopt the ratio $\Theta\approx 200$ giving $T_E\approx 0.25$. Therefore significantly more of the flare energy might penetrate to the chromosphere than to the photosphere. Furthermore, since the chromospheric plasma is less dense than the photospheric plasma, the wave energy would be able to change the physical conditions, heating the plasma and moving the magnetic field, in the chromosphere more than in the photosphere. It would therefore take less energy to brighten chromospheric plasma to UV temperatures than it would to produce a photospheric field change.

The observations are consistent with the following picture. An Alfv\'{e}n wave from the field reconnection site in the corona propagates field changes outward in all directions near the onset of the impulsive phase (see Cliver and Petrie~2012), including downwards through the chromosphere and into the photosphere, causing the photospheric field changes. These field changes are abrupt and permanent, and are complete within ten minutes or so. In contrast, the chromosphere, being subject to more energy from the corona and being populated by less dense plasma, emits in UV in the form of flare kernels, ribbons and sequential chromospheric brightenings during all phases of the flare.


\acknowledgements{B.M.J. carried out this work through the National Solar Observatory Research Experiences for Undergraduate (REU) site program, which is cofunded by the Department of Defense in partnership with the National Science Foundation REU Program. This work utilizes data obtained by the Global Oscillation Network Group (GONG) program, managed by the National Solar Observatory, which is operated by AURA, Inc. under a cooperative agreement with the National Science Foundation. The data were acquired by instruments operated by the Big Bear Solar Observatory, High Altitude Observatory, Learmonth Solar Observatory, Udaipur Solar Observatory, Instituto de Astrof\'{\i}sica de Canarias, and Cerro Tololo Interamerican Observatory. This work uses data from the TRACE mission. \textit{Transition Region and Coronal Explorer} (TRACE) is a mission of the Stanford-Lockheed Institute for Space Research (a joint program of the Lockheed-Martin Advanced Technology Center's Solar and Astrophysics Laboratory and Stanford's Solar Observatories Group), and part of the NASA Small Explorer Program.}

\end{document}